\documentclass[aps,prl,onecolumn,superscriptaddress,groupedaddress]{revtex4}  
\usepackage{graphicx}  
\usepackage{color, soul}
\usepackage{xcolor}
\usepackage{dcolumn}   
\usepackage{bm}        
\usepackage{amssymb}   
\usepackage{changepage}
\usepackage{amsmath}
\usepackage{hyperref}

\usepackage{graphicx}  
\usepackage{caption}
\usepackage{setspace}
\usepackage{float}
\usepackage{blindtext}
\hypersetup{
  colorlinks=false,
  linkbordercolor=blue
}

\usepackage[a4paper,bindingoffset=0.1in,%
            left=1.5cm,right=1.5cm,top=2cm,bottom=2.5cm,%
            footskip=0.5in]{geometry}
\usepackage{blindtext}

\begin{document}


\newlength{\halfpagewidth}
\setlength{\halfpagewidth}{\linewidth}
\divide\halfpagewidth by 2
\newcommand{\leftsep}{%
\noindent\raisebox{4mm}[0ex][0ex]{%
\makebox[\halfpagewidth]{\hrulefill}\hbox{\vrule height 3pt}}%
}
\newcommand{\rightsep}{%
\noindent\hspace*{\halfpagewidth}%
\rlap{\raisebox{-3pt}[0ex][0ex]{\hbox{\vrule height 3pt}}}%
\makebox[\halfpagewidth]{\hrulefill} } 
    
\title{Atomistic Simulation of Phonon and Magnon Thermal Transport across the Ferro-Paramagnetic Transition}
\author{Yanguang~Zhou}
\affiliation{Mechanical and Aerospace Engineering Department, University of California Los Angeles, Los Angeles, CA 90095, USA}

\author{Julien~Tranchida}
\email{jtranch@sandia.gov}
\affiliation{Multiscale Science Department, Sandia National Laboratories, P.O. Box 5800, MS 1322, Albuquerque, NM 87185, USA}
    
\author{Yijun~Ge}
\affiliation{Mechanical and Aerospace Engineering Department, University of California Los Angeles, Los Angeles, CA 90095, USA}        

\author{Jayathi~Murthy}
\email{jmurthy@ucla.edu}
\affiliation{Mechanical and Aerospace Engineering Department, University of California Los Angeles, Los Angeles, CA 90095, USA}

\author{Timothy~S.~Fisher}
\email{tsfisher@ucla.edu}
\affiliation{Mechanical and Aerospace Engineering Department, University of California Los Angeles, Los Angeles, CA 90095, USA}         

\date{\today}
\begin{spacing}{2}
\begin{abstract}
A temperature-dependent approach involving Green-Kubo equilibrium atomic and spin dynamics (GKEASD) is reported to assess phonon and magnon thermal transport processes accounting for phonon-magnon interactions. Using body-center cubic (BCC) iron as a case study, GKEASD successfully reproduces its characteristic temperature-dependent spiral and lattice thermal conductivities. The non-electronic thermal conductivity, i.e., the sum of phonon and magnon thermal conductivities, calculated using GKEASD for BCC Fe agrees well with experimental measurements. Spectral energy analysis reveals that high-frequency phonon-magnon scattering rates are one order of magnitude larger than those at low frequencies due to energy scattering conservation rules and high densities of states. Higher temperatures further accentuate this phenomenon. This new framework fills existing gaps in simulating thermal transport across the ferro- to para-magnetic transition. Future application of this methodology to phonon- and magnon-dominant insulators and semiconductors will enhance understanding of emerging thermoelectric, spin caloritronic and superconducting materials. 
\end{abstract}
\setulcolor{blue}
\maketitle

\section{\label{sec:level1}I. INTRODUCTION}

A better understanding of heat transfer considering interactions among different temperature-induced excitations in crystals, e.g., phonons, electrons and spins, is of great importance in many disciplines, including thermoelectric \cite{Uchida2010, Flipse2012, Uchida2011}, spin caloritronic \cite{Bauer2012} and superconducting \cite{Chen1988} materials. Unlike phonons and electrons, whose thermal transport properties have been well studied, the heat transport behavior of magnons - collective excitations of magnetic spins - is poorly understood at the fundamental level, and little is known about the influence of phonon-magnon and magnon-magnon scattering on heat transfer. For example, the experimental lattice thermal conductivities of magnetic materials such as body-centered cubic (BCC) iron \cite{Fulkerson1966, Backlund1961}, face-centered cubic (FCC) nickel \cite{Powell1965}, YMnO$_\text{3}$, LuMnO$_\text{3}$ and ScMnO \cite{Sharma2004}, as well as CrN \cite{Tomes2011, Jankovsky2014}, show significantly different temperature dependences near the Curie temperature ($T_\text{c}$), compared to the typical $1/T$ relation at high temperatures when only phonons are considered \cite{Holland1963, Callaway1959, Zhou2018}. The magnetic configuration changes from a ferromagnetic state at low temperature, in which spins are aligned to a paramagnetic state above the Curie temperature, in which the spin configuration is disordered \cite{Kormann2014}. 

To model the thermal transport properties of magnetic materials, temperature-dependent lattice and magnetic excitations must be taken into consideration when calculating the inputs, e.g., heat current or force constants, for thermal conductivity. To date, theoretical studies have sought to develop explanations for mutual interactions between phonons and magnons \cite{Kormann2014, Stockem2018, Sabiryanov1999, Liao2014, Liu2017}. However, most such studies provide only a partial treatment based on thermodynamic properties \cite{Kormann2014, Dick2012}, average phonon-magnon relaxation times \cite{Stockem2018}, and phonon-magnon temperature gradients \cite{Liao2014}. A robust framework to compute thermal transport properties such as thermal conductivity and modal scattering rates in magnetic materials is still lacking.

In this paper, a temperature-dependent method - Green-Kubo equilibrium atomic and spin dynamics (GKEASD) - based on linear response theory \cite{Kubo1957} and spin-lattice dynamics \cite{Tranchida2018, Ma2016} is reported to calculate the thermal transport properties of phonons and magnons in magnetic materials. In GKEASD, phonon-phonon scattering, phonon-magnon scattering and magnon-magnon scattering are inherently included. As a proof of concept, our study focuses on a model system consisting of a simple transition-metal ferromagnet, BCC iron, with a Curie temperature of 1043 K. Using this methodology, we successfully reproduce the characteristic temperature-dependent non-electronic thermal conductivity observed in experiments for magnetic materials \cite{Fulkerson1966, Backlund1961, Powell1965, Sharma2004, Tomes2011, Quintela2009}. Modal level phonon-phonon, magnon-magnon and phonon-magnon scattering rates are then quantified using spectral energy density analysis. The agreement between theoretical predictions and experimental measurements establishes the reliability of the methodology.

\section{\label{sec:level1}II. Computational methods}

\subsection{\label{sec:level2}A. Spin Lattice Dynamics}
\begin{figure*}
\setlength{\abovecaptionskip}{0.1in}
\setlength{\belowcaptionskip}{-0.1in}
\includegraphics [width=6.5in]{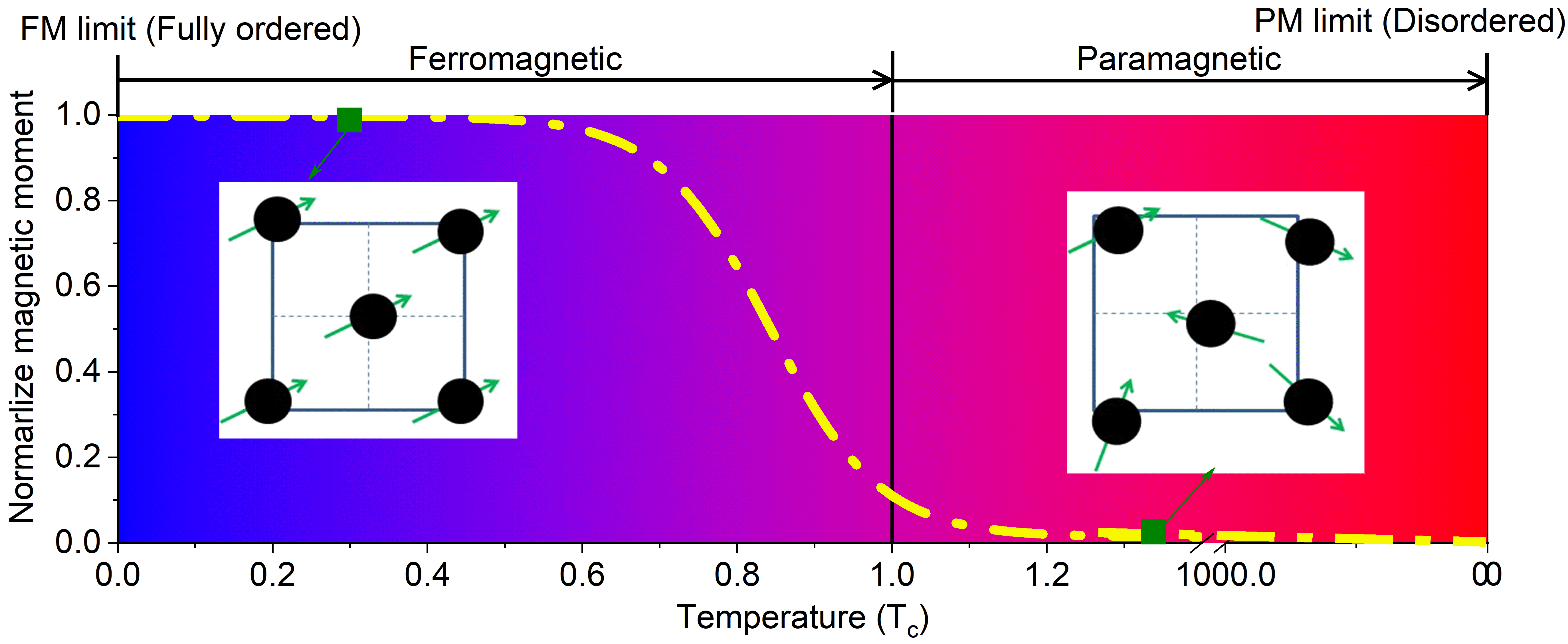}
\caption{Magnetic state of the system with no external magnetic field as a function of temperature. The arrows indicate notional spin directions in the system.}
\label{fig:F1}
\end{figure*}
Here, a symplectic and scalable algorithm for spin lattice dynamics that was recently developed \cite{Tranchida2018, Ma2016} and embedded in the Large-scale Atomic/Molecular Massively Parallel Simulator (LAMMPS) \cite{Plimpton1995} is applied to describe the atomic spins in magnetic crystals from the ferromagnetic limit to the paramagnetic limit (\textbf{Figure ~\ref{fig:F1}}). Compared to classical molecular dynamics, this new algorithm augments the phase space by adding a classical spin vector $\vec{s}$ to each magnetic atom $i$, in addition to its position $\vec{r}$ and momentum $\vec{p}$. The motion equations of atoms and spins can be written as
\begin{small}
\begin{equation}
\frac{d{{{\vec{r}}}_{i}}}{dt}=\frac{{{{\vec{p}}}_{i}}}{{{m}_{i}}}
\label{eqn:A1}  
\end{equation}
\end{small}
\begin{small}
\begin{equation}
\frac{d{{{\vec{p}}}_{i}}}{dt}=\sum\limits_{i\ne j}^{N}{\left[ -\frac{dV(|{{{\vec{r}}}_{ij}}|)}{d|{{{\vec{r}}}_{ij}}|}+\frac{dJ(|{{{\vec{r}}}_{ij}}|)}{d|{{{\vec{r}}}_{ij}}|}{{{\vec{s}}}_{i}}\cdot {{{\vec{s}}}_{j}} \right]{{{\vec{e}}}_{ij}}}\
\label{eqn:A2}  
\end{equation}
\end{small}
\begin{small}
\begin{equation}
\frac{d{{{\vec{s}}}_{i}}}{dt}={{\vec{f}}_{i}}\times {{\vec{s}}_{i}}
\label{eqn:A3}  
\end{equation}
\end{small}

\noindent
where ${{\vec{e}}_{ij}}$ is the unit vector along ${{\vec{r}}_{ij}}$, $\vec{f_{i}}$ is the analog of a spin force applied on the spin. $J(|{{{\vec{r}}}_{ij}}|)$ is the magnetic coupling exchange constant, which originates from two main contributions: (i) direct ferromagnetic exchange between the orbitals localized on centers of ions $i$ and $j$, and (ii) spin and charge polarizations effects carried by nonmagnetic orbitals. Spin dynamics simulations without lattice vibrations, i.e., fixed atomic position, are implemented in a $NVT$ ensemble (constant number of particles, volume and temperature) for 200 ps to reach the target temperature, and then a $NVE$ ensemble (constant number of particles, volume and energy) is run for 200 ps to compute the heat flux associated with the spins. In the $NVT$ simulations, the thermal fluctuations in the magnetic system are described following the Langevin approach \cite{Tranchida2018, Antropov1995, Carcia1998}.

For a system with lattice vibrations, a $NVT$ simulation for 200 ps is first run to reach the target temperature. Then, a $NVE$ ensemble for 200 ps is used to obtain the heat flux fluctuation of the magnons. A time step of 0.2 fs is used in all simulations. The spin potential used in our study is fitted using first-principles calculations (details can be found in \cite{Tranchida2018}). The magnon dispersion generated from this potential fits experiments well (see details below). Mechanical interactions among atoms are computed using the embedded-atom method potential \cite{Chamati2006}, and the interactions between spins is modeled with the Bethe-Slater curve \cite{Yosida1996}
\begin{equation}
J\left( {|{r}_{ij}|} \right)=4\alpha {{\left( \frac{{|{r}_{ij}|}}{\delta } \right)}^{2}}\left( 1-\gamma {{\left( \frac{{|{r}_{ij}|}}{\delta } \right)}^{2}} \right){{e}^{-{{\left( \frac{{|{r}_{ij}|}}{\delta } \right)}^{2}}}}\Theta (R-{|{r}_{ij}|})
\label{eqn:A4}  
\end{equation}
where the fitting parameters $\alpha$, $\gamma$ and $\delta$ are 25.498 meV, 0.281 and 1.999 $\text{\AA}$, respectively based on the $J$ values for BCC iron \cite{Pajda2001}. $\Theta (R-{{r}_{ij}})$ is the Heaviside step function, and $R$ is the cutoff radius, which is 4 $\text{\AA}$ in our study. 

\subsection{\label{sec:level2}B. First Principles Calculations}
In this paper, all first-principles simulations are implemented via Vienna Ab-initio Simulation Package (VASP) based on density functional theory. The pseudopotential with a generalized gradient approximation parameterized by Perdew-Burke-Ernzerhof {theory \cite{Perdew1986} for the exchange-correlation functional} is used to depict the system. Periodic boundary conditions are applied in the three directions. A plane wave basis with a cutoff energy of 520 eV is used in all simulations. The Monkhorst-Pack scheme is used to generate 2$\times$2$\times$2 k-point mesh for the primitive cell and 6$\times$6$\times$6 supercell. Before performing electrostatic potential or interatomic force constant calculations, the atomic structure and cell size are fully relaxed until the energy difference and the Hellman-Feynman force converge to within 1$\times10^{-6}$ eV and 1$\times10^{-5}$ eV/\AA, respectively. The 6$\times$6$\times$6 supercell with an energy criterion of 1$\times10^{-6}$ eV in the self consistent calculation is used to extract 2$^{nd}$ and 3$^{rd}$ order force constants.

\subsection{\label{sec:level2}C. Green-Kubo Equilibrium Atomic and Spin Dynamic Simulations}

\begin{figure}[H]
\begin{center}
\setlength{\abovecaptionskip}{0.1in}
\setlength{\belowcaptionskip}{-0.2in}
\includegraphics [width=6.5in]{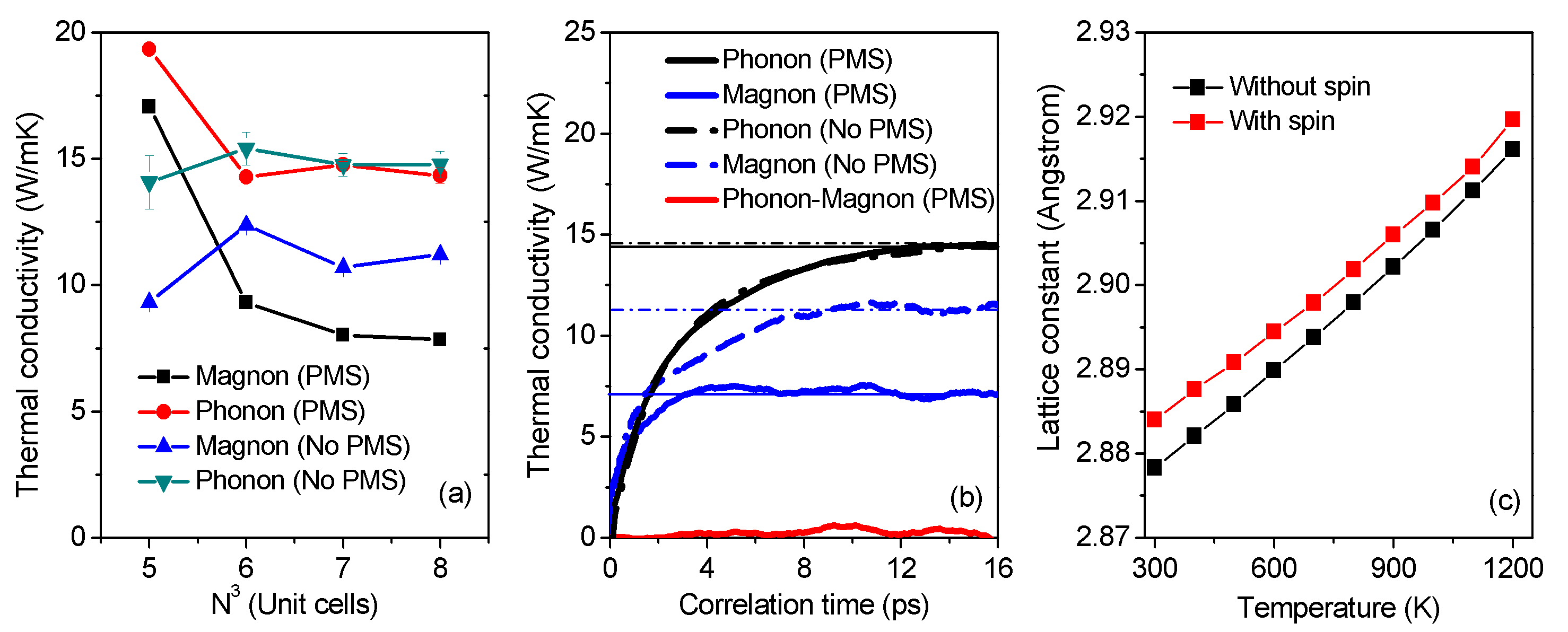}
\caption{(a) Convergence test for GKESAD in BCC iron; N is the number of unit cells along one direction. Computations are performed with and without phonon-magnon scattering (PMS). (b) Thermal conductivity of phonons and magnons in BCC iron calculated using GKEASD at 300 K. (c) Lattice constant as a function of temperature with and without spin effects.}
\label{fig:F2}
\end{center}
\end{figure}

During molecular dynamics simulations for lattice thermal conductivity calculations, we first run the $NVT$ ensemble for 200 ps with a time step of 0.2 fs to allow the lattice and spin fields attain the target temperature. Then a canonical ensemble is run for the next 400 ps to generate heat current, which is the input for the calculation of lattice thermal conductivity using Green-Kubo equilibrium atomic and spin dynamics. For each case, 30 independent runs are performed to obtain a stable average thermal conductivity. The correlation time considered in our simulations is 16 ps, which is long enough to obtain a converged and steady thermal conductivity (\textbf{Figure ~\ref{fig:F2}a}). An 8$\times$8$\times$8 unit cell box is used in all equilibrium molecular dynamics simulations, for which size effects may be ignored (\textbf{Figure ~\ref{fig:F2}b}). Lattice expansion due to temperature and spin is considered in our results (\textbf{Figure ~\ref{fig:F2}c}).

For the magnon thermal conductivity computations, 30 independent runs were used to obtain a converged thermal conductivity. A domain of 8$\times$8$\times$8 unit cells was used, and the correlation time was 16 ps, which is long enough to obtain a converged and size-independent magnon thermal conductivity (\textbf{Figure ~\ref{fig:F2}b}).

\subsection{\label{sec:level2}D. Validation of Classical Mechanical Potential}
\begin{figure}[H]
\begin{center}
\setlength{\abovecaptionskip}{0.1in}
\setlength{\belowcaptionskip}{-0.1in}
\includegraphics [width=6.5in]{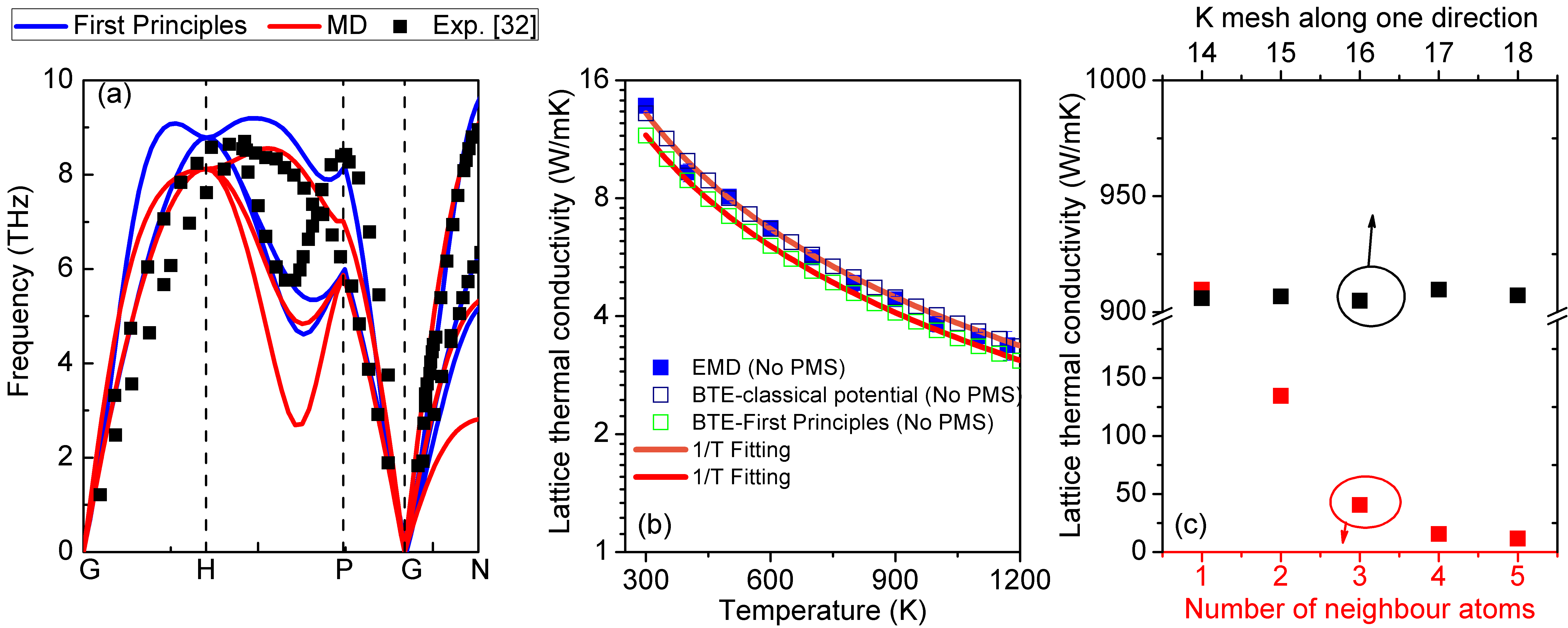}
\caption{(a) Phonon dispersion from molecular dynamics and first-principles calculations. The black dots are experimental values \cite{Minkiewicz1967}. (b) The temperature-dependent lattice thermal conductivity of BCC iron without the influence of spins. (c) Convergence test for Boltzmann transport equation calculations, N is the number of the neighbour and K is the mesh of the wave vector along one direction.}
\label{fig:F3}
\end{center}
\end{figure}
To assess the accuracy of the potential, we compare the phonon dispersion (\textbf{Figure ~\ref{fig:F3}a}) as well as lattice thermal conductivity (\textbf{Figure ~\ref{fig:F3}b}) from molecular dynamics and $ab$ $initio$ calculations. Both the phonon dispersion and the lattice thermal conductivity computed using a classical embedded-atom method potential agree with those calculated using a first-principles approach somehow. The $ab$ $initio$ thermal conductivity is calculated via the Boltzmann transport equation which can be solved by expanding the scattering term into its first-order perturbation $n^{1}$. Without considering impurities and boundaries, the linearized Boltzmann transport equation may be recast as \cite{Li2014}
\begin{equation}
\begin{aligned}
-c(\vec{q},\nu )\frac{\partial \bar{n}(\vec{q},\nu )}{\partial T}=&\sum\limits_{({\vec{q}}',{\nu }'),\ ({\vec{q}}'',{\nu }'')}{\left[ \Gamma _{({\vec{q}}',{\nu }'),\ ({\vec{q}}'',{\nu }'')}^{(\vec{q},\nu )}(n_{(\vec{q},\nu )}^{1}+n_{({\vec{q}}',{\nu }')}^{1}-n_{({\vec{q}}'',{\nu }'')}^{1})+ \right.} \\ 
 & \ \ \ \ \ \ \ \ \ \ \ \ \ \ \ \ \ \frac{1}{2}\left. \Gamma _{(\vec{q},\nu )}^{({\vec{q}}',{\nu }'),\ ({\vec{q}}'',{\nu }'')}(n_{(\vec{q},\nu )}^{1}-n_{({\vec{q}}',{\nu }')}^{1}-n_{({\vec{q}}'',{\nu }'')}^{1}) \right] \\ 
\end{aligned}
\label{eqn:A5}  
\end{equation}
\noindent
where $(\vec{q},\nu )$ is the phonon mode with wave vector $\vec{q}$ and branch $\nu$. $c$, $\bar{n}$ and $T$ are the specific heat capacity, equilibrium phonon population and system temperature, respectively. $\Gamma$ is the scattering rate at equilibrium of a process where two phonons combine to generate a third phonon or when one phonon splits into two phonons. The scattering rate matrix can be obtained using Fermi\rq s Golden Rule. The lattice thermal conductivity may be found by computing the heat flux from the computed phonon distribution and using Fourier\rq s law. The wave vector mesh density is varied from 14$\times$14$\times$14 to 17$\times$17$\times$17, and the neighbor cutoffs range from 1$^{st}$ to 5$^{th}$. The latter are used to compute the 3$^{rd}$ order force constant which is key parameter to obtain the lattice thermal conductivity in Boltzmann transport equation. Based on these variations, a wave vector mesh of 17$\times$17$\times$17 with a 5$^{th}$ neighbour cutoff was chosen in our all Boltzmann transport equation calculations to produce a converged lattice thermal conductivity (\textbf{Figure ~\ref{fig:F3}c}).

\section{\label{sec:level1}III. Theory}
\subsection{\label{sec:level2}A. Thermal Conductivity based on Heat Flux Fluctuation}
\begin{figure}
\includegraphics [width=3.4in]{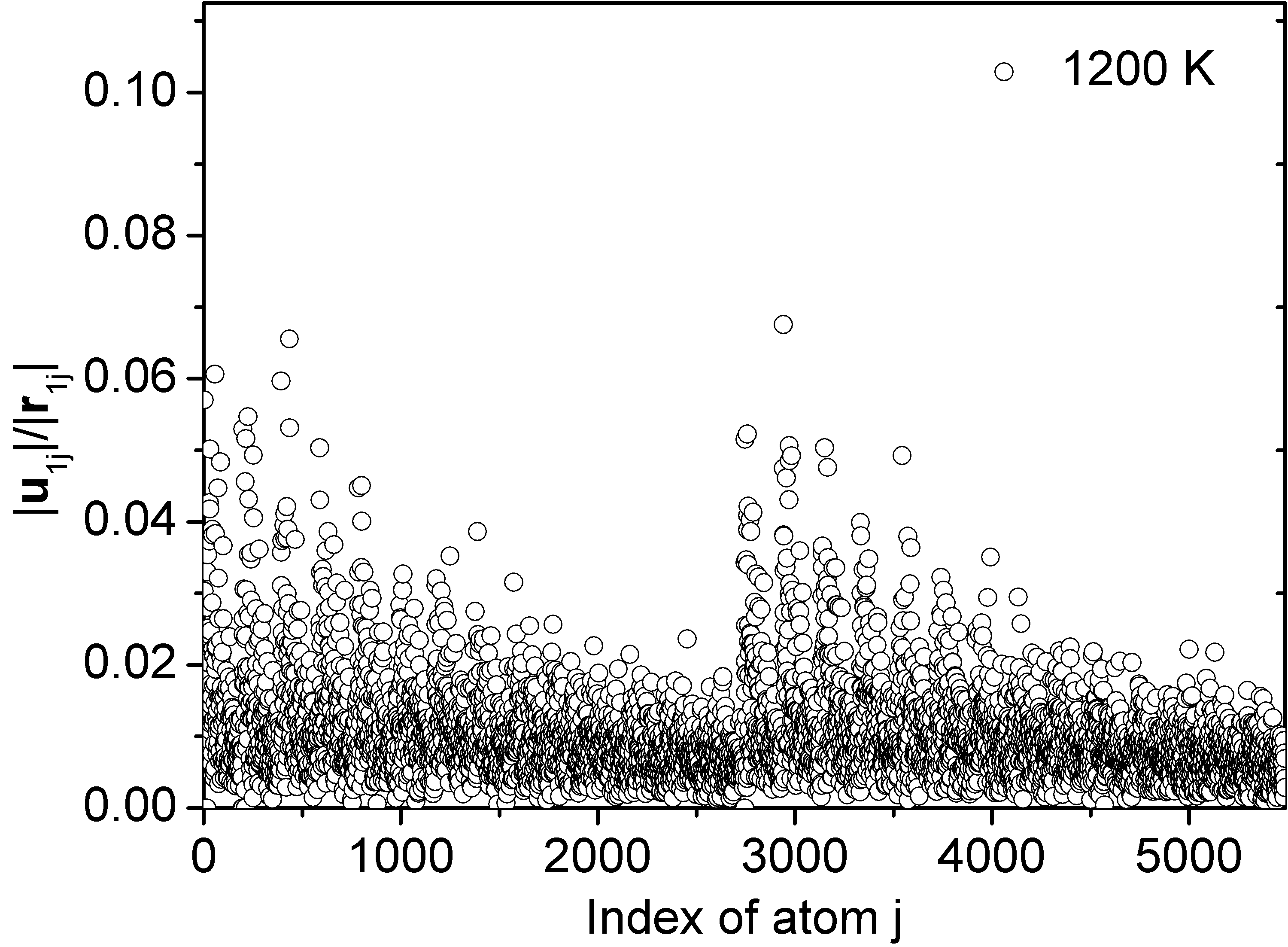}
\caption{Relative displacement between atoms $i$ and $j$.}
\label{fig:F4}
\end{figure}

The energy $E$ of magnetic crystals must account for terms coupling the magnetic spins ($E_{spin}$) to the lattice ($E_{kinetic}$ and $E_{potential}$) through the following expression \cite{Tranchida2018}:
\begin{equation}
\begin{aligned}
&E = {{E}_{kinetic}}+{{E}_{potential}}+{{E}_{spin}} \\ 
&=\sum\limits_{i=1}^{N}{\frac{|\vec{p}{{|}^{2}}}{2{{m}_{i}}}+\sum\limits_{i,\ j=1}^{N}{V({{{\vec{r}}}_{ij}})}-\sum\limits_{i,\ j,\ i\ne j}^{N}{J(|{{{\vec{r}}}_{ij}}|){{{\vec{s}}}_{i}}\cdot {{{\vec{s}}}_{j}}}} \\ 
\end{aligned}
\label{eqn:A6}  
\end{equation}
in which $\vec{r}$, $\vec{p}$ and $\vec{s}$ are the position, momentum and spin vectors of the atoms, respectively; the negative sign on the last term indicates that the ground state energy of the system considering spin becomes lower. Phonons and magnons are coupled in Eq.\ (\ref{eqn:A6}) via atomistic positions, i.e., $\vec{r}_{ij}$. The heat flux due to lattice vibrations takes the form \cite{Hardy1963}:

\begin{equation}
\begin{aligned}
 & Q_{lattice}^{''x}=\frac{1}{V}\sum\limits_{i}^{{}}{\frac{d(E_{i}^{kinetic}+E_{i}^{potential})r_{i}^{x}}{dt}} \\ 
 & =\frac{1}{V}\left\{ \sum\limits_{i}^{{}}{{{e}_{i}}\cdot v_{i}^{x}}+\frac{1}{2}\sum\limits_{i=1}{\sum\limits_{j=1;\ j\ne i}^{{}}{\left( {{{\vec{F}}}_{ij}}\cdot {{{\vec{v}}}_{i}} \right)\cdot r_{ij}^{x}}} \right\} \\ 
\end{aligned}
\label{eqn:A7}  
\end{equation}
\noindent
where ${{e}_{i}}$ and ${{\vec{v}}_{i}}$ are the energy and the velocity of atom $i$, respectively. ${{\vec{F}}_{ij}}$ represents the force between two atoms, $V$ and $S$ are the volume and the cross section area of the system, respectively.  and the heat flux associated with spin is expressed as
\begin{equation}
\begin{aligned}
 {{Q}^{''}_{spin, ij}}=&\frac{1}{S}\frac{dE_{ij}^{spin}}{dt} \\ 
  =&\frac{1}{S}J(|{{{\vec{r}}}_{ij}}|)\left( {{{\vec{s}}}_{i}}\cdot \frac{d{{{\vec{s}}}_{j}}}{dt}+{{{\vec{s}}}_{j}}\cdot \frac{d{{{\vec{s}}}_{i}}}{dt} \right)+\\
&\frac{1}{S}\frac{dJ(|{{{\vec{r}}}_{ij}}|)}{dt}{{{\vec{s}}}_{i}}\cdot {{{\vec{s}}}_{j}} \\ 
\end{aligned}
\label{eqn:A8}  
\end{equation}
where $|{{\vec{r}}_{ij}}|$ is the distance between atoms $i$ and $j$:
\begin{equation}
\begin{aligned}
{{\vec{r}}_{ij}}&={{\vec{r}}_{i}}-{{\vec{r}}_{j}}\\ 
&=[\vec{r}_{i}^{0}+{{\vec{u}}_{i}}(t)]-[\vec{r}_{j}^{0}+{{\vec{u}}_{j}}(t)]\\ 
&=\vec{r}_{ij}^{0}+{{\vec{u}}_{ij}}(t)
\label{eqn:A9}  
\end{aligned}
\end{equation}
in which, $\vec{r}_{{}}^{0}$ and $\vec{u}$ are the equilibrium position and displacement of atoms, respectively. Since $|\vec{r}_{ij}^{0}|\gg |{{\vec{u}}_{ij}}|$ (\textbf{Figure ~\ref{fig:F4}}), we assume that ${{\vec{r}}_{ij}}$ is time independent. Therefore, Eq. (\ref{eqn:A9}) can be rewritten in the form 
\begin{equation}
\begin{aligned}
{{Q}^{''}_{spin, ij}}&=\frac{1}{S}\frac{dE_{ij}^{spin}}{dt}\\ 
&\approx \frac{J(|{{\vec{r}}_{ij}}|)}{S}\left( {{{\vec{s}}}_{i}}\cdot \frac{d{{{\vec{s}}}_{j}}}{dt}+{{{\vec{s}}}_{j}}\cdot \frac{d{{{\vec{s}}}_{i}}}{dt} \right)
\end{aligned}
\label{eqn:A10}  
\end{equation}

At the same time, if we assume that atoms $i$ and $j$ are separated by an imaginary interface, the heat flux across the interface can be expressed as
\begin{equation}
\begin{aligned}
Q^{''}_{spin}&=\frac{1}{2S}\sum\limits_{i\in L}{\sum\limits_{i\in R}{{{Q^{''}}_{spin, ij}}}}\\ 
&=\frac{1}{2S}\sum\limits_{i\in L}{\sum\limits_{i\in R}{J(|{{{\vec{r}}}_{ij}}|)\left( {{{\vec{s}}}_{i}}\cdot \frac{d{{{\vec{s}}}_{j}}}{dt}+{{{\vec{s}}}_{j}}\cdot \frac{d{{{\vec{s}}}_{i}}}{dt} \right)}}\\ 
&=\frac{1}{S}\sum\limits_{i\in L}{\sum\limits_{i\in R}{J(|{{{\vec{r}}}_{ij}}|)\frac{d{{{\vec{s}}}_{j}}}{dt}\cdot {{{\vec{s}}}_{i}}}}
\label{eqn:A11} 
\end{aligned} 
\end{equation}
where $L$ and $R$ indicate left and right sides of the imaginary interface, respectively. The factor 1/2 addresses the ergodicity of the system. In addition, for spin lattice dynamics systems, 
\begin{equation}
\begin{aligned}
\frac{d{{{\vec{s}}}_{i}}}{dt}&={{\vec{f}}_{i}}\times {{\vec{s}}_{i}}=-\frac{1}{\hbar }\frac{\partial {{E}_{mag}}}{\partial {{{\vec{s}}}_{i}}}\times {{\vec{s}}_{i}}
\label{eqn:A12}  
\end{aligned}
\end{equation}
in which ${{\vec{f}}_{i}}$ is the analog of a force applied on the spin. Finally, the heat flux across the imaginary interface can be written as
\begin{equation}
 Q_{spin}^{''x} =-\frac{1}{S\hbar}\sum\limits_{i\in L}{\sum\limits_{i\in R}{J(|{{{\vec{r}}}_{ij}}|)\left( \frac{\partial {{E}_{mag}}}{\partial {{{\vec{s}}}_{j}}}\times {{{\vec{s}}}_{j}} \right)\cdot {{{\vec{s}}}_{i}}}}
\label{eqn:A13}  
\end{equation}

\noindent
Based on linear response theory \cite{Kubo1957}, Eq.\ (\ref{eqn:A7}) and Eq.\ (\ref{eqn:A13}), the thermal conductivity of a magnetic system can be divided into contributions from lattice vibrations, ${{\kappa }_{phonon}}$, spin-related fluctuations, ${{\kappa }_{magnon}}$, and a term resulting from lattice-spin interactions, ${{\kappa }_{cross}}$:
\begin{equation}
\begin{aligned}
&\kappa ={{\kappa }_{phonon}}+{{\kappa }_{magnon}}+{{\kappa }_{cross}} \\ 
&=\frac{V}{{{k}_{b}}{{T}^{2}}}\int{\left[ \underbrace{\left\langle Q_{lattice}^{''x}(t)\cdot Q_{lattice}^{''x}(0) \right\rangle }_{phonon} \right.}\\ 
&+\underbrace{\left\langle Q_{spin}^{''x}(t)\cdot Q_{spin}^{''x}(0) \right\rangle }_{magnon}+\left. \underbrace{2\left\langle Q_{spin}^{''x}(t)\cdot Q_{lattice}^{''x}(0) \right\rangle }_{cross} \right]dt \\ 
\end{aligned}
\label{eqn:A14}  
\end{equation}
\noindent
where ${{k}_{b}}$ is the Boltzmann constant, and $T$ is the temperature of the system. We note that ${{\kappa }_{cross}}$ is not the result of phonon-magnon scattering, but of the cross-correlation between the phonon and magnon heat fluxes. Physically, the cross term represents the interaction between the heat carried by the phonons and the heat carried by the magnons that can alter pure phonon and spin heat flow. The effect of magnon scattering on phonon transport is elucidated by calculating the phonon thermal conductivity separately using ($E={{E}_{kinetic}}+{{E}_{potential}}+{{E}_{spin}}$) and ($E={{E}_{kinetic}}+{{E}_{potential}}$) and comparing the two values. Similarly, the influence of phonon scattering on magnon thermal conductivity is assessed by computing the magnon thermal conductivity using ($E={{E}_{kinetic}}+{{E}_{potential}}+{{E}_{spin}}$) and ($E={{E}_{spin}}$) and comparing the two values. 

\subsection{\label{sec:level2}B. Spectral Energy Density}
The atomistic velocity and spin change frequency used for the spectral energy density calculations were output every 4 fs over a total sampling time of 80 ps. All the reduced results were averaged across 3 runs and using two time intervals of 40 ps each. A Gauss window of 0.1 THz was used to filter noise in the original data. 
\begin{figure}[H]
\begin{center}
\setlength{\abovecaptionskip}{0.1in}
\setlength{\belowcaptionskip}{-0.1in}
\includegraphics [width=5.5in]{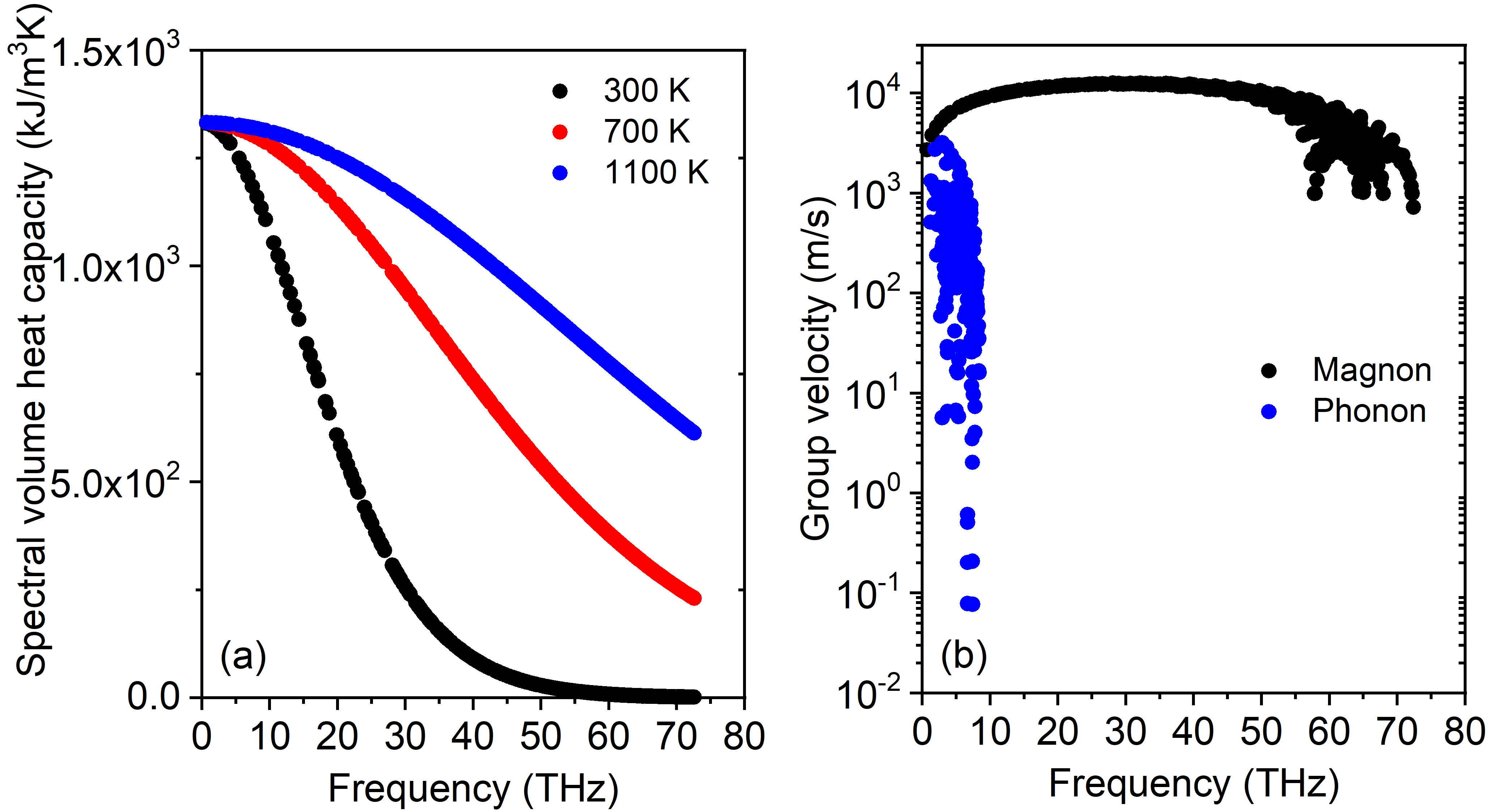}
\caption{(a) Spectral volumetric heat capacity of phonons and magnons and (b) group velocity of phonons and magnons. Phonons and magnons are both bosons, therefore phonons and magnons with the same frequency have the same spectral volumetric heat capacity.}
\label{fig:F5}
\end{center}
\end{figure}
The eigenvectors and frequencies of the phonon mode $(\vec{q},\nu )$ and magnon mode $(\vec{k})$ used in our simulations are computed by solving the phonon dynamical matrix equation
\begin{equation}
\omega (\vec{q},\nu )\vec{e}(\vec{q},\nu )=\omega (\vec{q},\nu )\text{D}(\vec{q})
\label{eqn:A15}  
\end{equation}
and the magnon dynamical matrix equation
\begin{equation}
\omega (\vec{k})\vec{e}(\vec{k})=\text{D}(\vec{k})\vec{e}(\vec{k})
\label{eqn:A16}  
\end{equation}
where $\text{D}(\vec{q})$ and $\text{D}(\vec{k})$ are the lattice and spin dynamical matrices
\begin{equation}
\text{D}(\vec{q})=\frac{1}{m}\sum\limits_{l}{{K}(0,\ l)\exp (i\vec{k}[\vec{r}(l)-\vec{r}(0)])}
\label{eqn:A17}  
\end{equation}
and
\begin{equation}
\text{D}(\vec{k})=\frac{1}{\hbar }\sum\limits_{l}{J(0,\ l)}\cdot \left[ 1-\exp (i\vec{k}[\vec{r}(l)-\vec{r}(0)]) \right]
\label{eqn:A18}  
\end{equation}
respectively. Here, ${K}$ is the force constant matrix and $J(0,\ l)$ is magnetic exchange constant matrix. We note that Eq. (\ref{eqn:A18}) is derived assuming that the material is ferromagnetic, in which all spins have the same magnitude in one direction (only under this assumption can the spin motion equation be reduced to {Eq. (\ref{eqn:A16}) \cite{Kittel1987}}). The volumetric heat capacity $C_{V}$ and group velocity $v_{g}$ of phonons and magnons are shown in \textbf{Figure~\ref{fig:F5}}. Together with relaxation time $\tau$, the phonon or magnon thermal conductivity ${\kappa }_{ph\ or\ mag}$ can be obtained via
\begin{equation}
{{\kappa }_{ph\ or\ mag}}=\sum\limits_{(\vec{q},\ \nu )\ or\ (\vec{k})}{C_{V}}v_{g}^{2}\tau
\label{eqn:A19}  
\end{equation}
where $\tau = 1/2\Delta $. The linewidth $\Delta $, which is half the scattering rate $\Gamma$, can be calculated via spectral energy density analysis as:
\begin{equation}
{{\left| \Phi  \right|}^{2}}=\frac{{{I}_{p}}}{{{\left[ (\omega -{{\omega }_{p}}\ )/\Delta  \right]}^{2}}+1}
\label{eqn:A20}  
\end{equation}
where ${{I}_{p}}$ and ${{\omega }_{p}}$ are the magnitude and frequency at the peak center, respectively. $\Phi$ is the spectral energy, which takes the following form for phonons \cite{Thomas2010, Zhou2015}:
\begin{equation}
\begin{aligned}
 \Phi (\vec{q},\ \nu )\sim &\frac{1}{\sqrt{2\pi }{{t}_{0}}}\int_{0}^{{{t}_{0}}}{\sum\limits_{jl}{{{m}_{j}}\exp [i\vec{q}\cdot \vec{r}(jl)-i\omega t]\cdot }} \\ 
 & {{{\vec{e}}}^{*}}(j,\ \vec{q},\ \nu )\vec{v}(jl,\ t)dt \\ 
\end{aligned}
\label{eqn:A21}  
\end{equation}
\noindent

For magnons, the expression becomes \cite{Wu2018}:
\begin{equation}
\begin{aligned}
 \Phi (\vec{k},\ \mu )\sim &\frac{\hbar }{\sqrt{2\pi }{{t}_{0}}}\int_{0}^{{{t}_{0}}}{\sum\limits_{jl}{\exp [i\vec{k}\cdot \vec{r}(jl)-i\omega t]}}\cdot  \\ 
 & {{{\vec{e}}}^{*}}(j,\ \vec{k},\ \mu )\frac{d\vec{s}(jl,\ t)}{dt}dt \\ 
\end{aligned}
\label{eqn:A22}  
\end{equation}
\noindent
where $\vec{e}$ is the mode eigenvector of a phonon or magnon, and ${{t}_{0}}$ is the integration limit. 

\section{\label{sec:level1}IV. Results}
\subsection{\label{sec:level2}A. Non-electronic Thermal Conductivity}
\begin{figure}[H]
\begin{center}
\setlength{\abovecaptionskip}{0.1in}
\setlength{\belowcaptionskip}{-0.1in}
\includegraphics [width=5.5in]{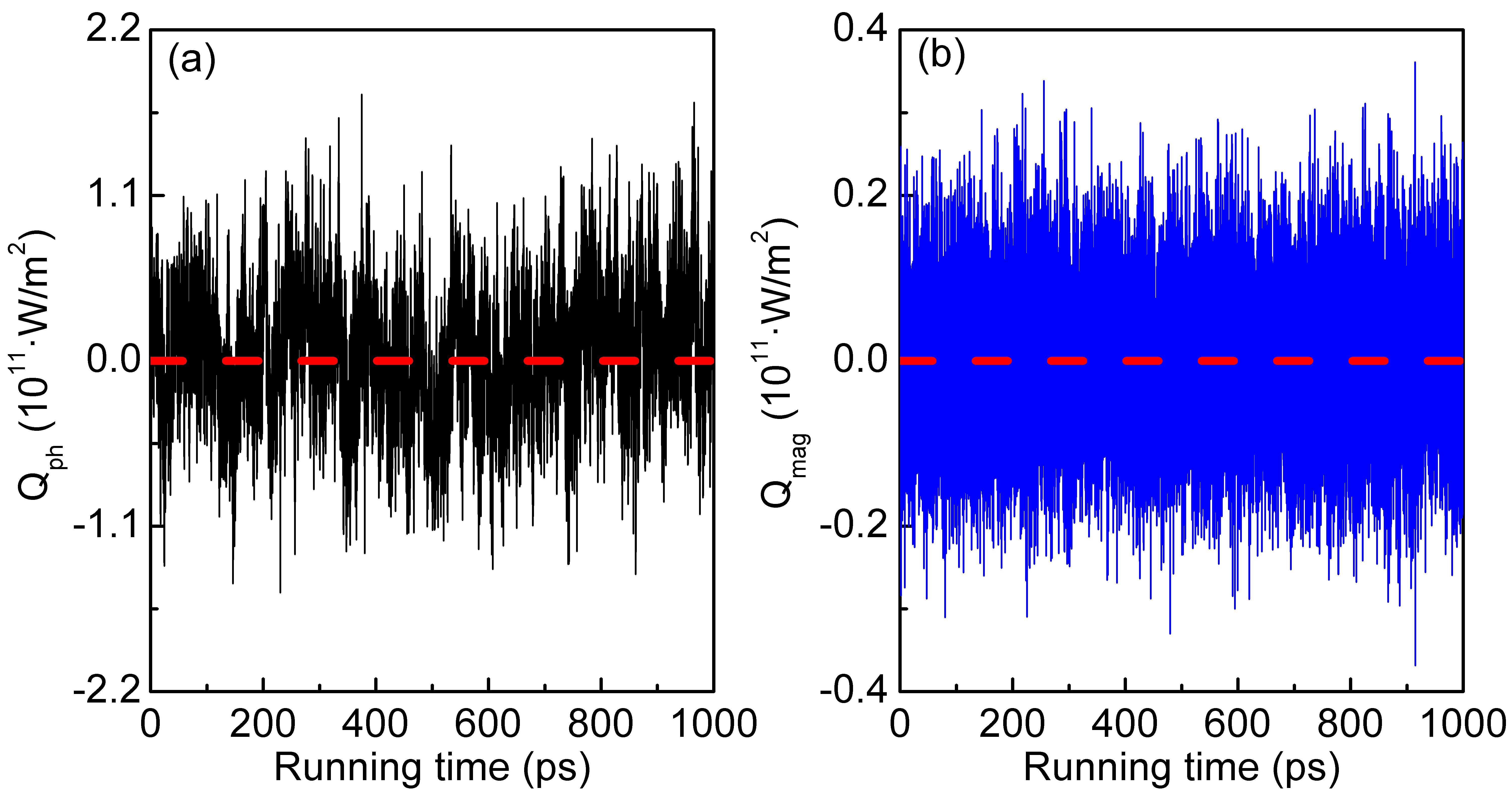}
\caption{Heat flux of (a) phonons and (b) magnons.}
\label{fig:F6}
\end{center}
\end{figure}
From Eqs.\ (\ref{eqn:A7}) and (\ref{eqn:A13}), the net heat flux of lattice vibrations ($Q_{lattice}^{''x}$) and spin fluctuations ($Q_{spin}^{''x}$) in EMD simulations should be zero, which is validated by our numerical results (\textbf{Figure ~\ref{fig:F6}a} and \textbf{\ref{fig:F6}b}). Furthermore, following linear response theory, the thermal conductivity of the two heat carriers should converge with increasing correlation time [Eq. (\ref{eqn:A14})], and this is also reflected in our simulations (\textbf{Figure ~\ref{fig:F2}b}).  

To evaluate thermal conductivity behavior at different temperatures, the order of the spin configuration in our system is calculated via
\begin{equation}
{{P}_{order}}=\sqrt{\sum\limits_{\alpha=x, y, z}{{{{\left( \sum\limits_{i}{s_{i,\ \alpha}^{T}} \right)}^{2}}}/{\sum\limits_{\alpha=x, y, z }{{{\left( \sum\limits_{i}{s_{i,\ \alpha }^{T=0}} \right)}^{2}}}}\;}}
\label{eqn:A23}  
\end{equation}
where ${{P}_{order}}=1$ in the ferromagnetic limit and ${{P}_{order}}=0$ in the paramagnetic limit. Small values of ${{P}_{order}}$ indicate that spins in the system are predominantly disordered, which leads to broken periodicity.

\begin{figure}[H]
\begin{center}
\setlength{\abovecaptionskip}{0.1in}
\setlength{\belowcaptionskip}{-0.1in}
\includegraphics [width=6.5in]{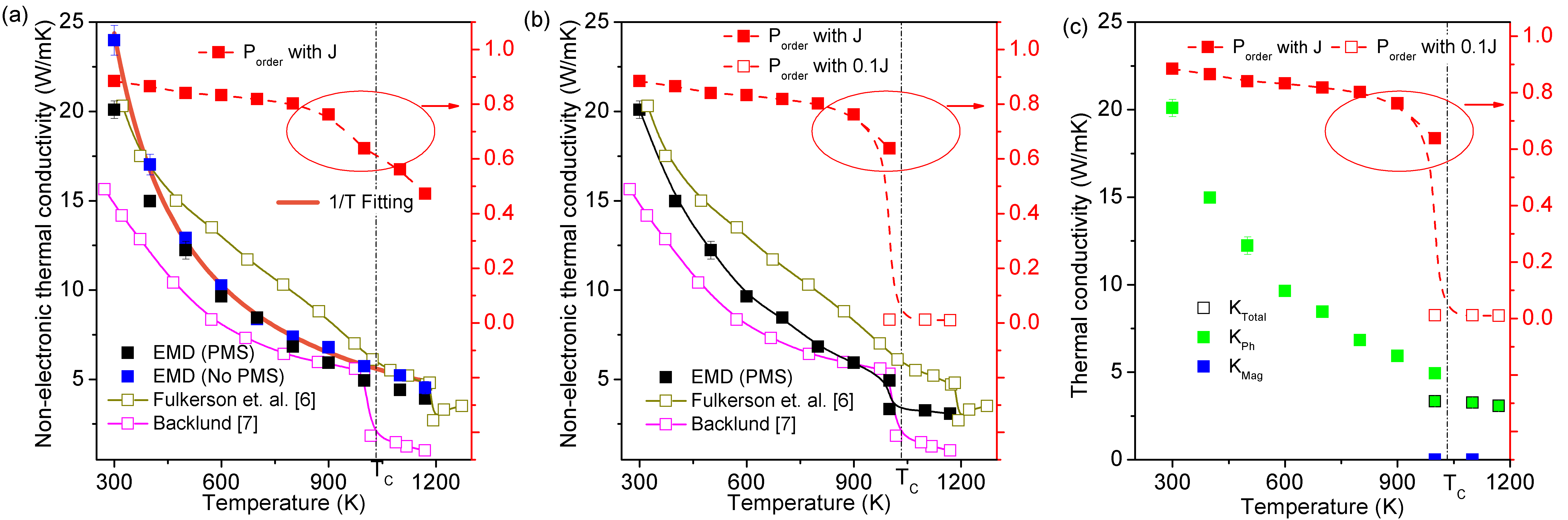}
\caption{(a) Non-electronic thermal conductivity for invariant magnetic exchange coefficient $J$, and (b) adjusted total thermal conductivity with a temperature-dependent magnetic exchange constant. The equilibrium molecular dynamics results are averaged over 30 independent runs. The experimental results are the so called \lq \lq lattice thermal conductivities\rq \rq  ~in Ref. \cite{Fulkerson1966, Backlund1961}. Here, the transition temperature for $J$ is  assumed to be 1043 K rather than the Curie temperature of 1200 K in order to better match the experimental data from Ref. \cite{Backlund1961}. (c) Thermal conductivity of phonons and magnons with and without phonon-magnon scattering. In (b) and (c) $J$ is assumed to vary as $J$ = $J_{0}$ for $T<T_{Curie}$, and $J = 0.1J_{0}$ for $T> = T_{Curie}$, with $J_{0}$ calculated using {Eq. (\ref{eqn:A4})}.}
\label{fig:F7}
\end{center}
\end{figure}

By considering the contributions of both phonons and magnons, the predicted total thermal conductivity is in broad agreement with experimental measurements \cite{Fulkerson1966, Backlund1961} over the temperature range from 300 K to 1200 K (\textbf{Figure ~\ref{fig:F7}a}) (the electrical thermal conductivity was subtracted from the experimental measurements). We note that the two experimental results \cite{Fulkerson1966, Backlund1961} in \textbf{Figure ~\ref{fig:F7}a} are themselves quite different from each other, and the disagreement between them is larger than that caused by phonon-magnon scattering in the simulations. Thus, they cannot definitively establish the accuracy of our phonon-magnon scattering models. A number of reasons may account for these differences: (i) loss of heat from the specimen due to conduction through leads, and (ii) impurities in the sample, which in Ref. \cite{Fulkerson1966} may be as high as 1\%. 

An important feature in \textbf{Figure ~\ref{fig:F7}a} is the sharp fall in thermal conductivity reported in the experimental data of Ref. \cite{Backlund1961} near the Curie temperature (around 1043 K) due to the ferromagnetic-to-paramagnetic transition. Previously published atomic spin dynamics or spin lattice dynamics simulations have been known to smooth the ferromagnetic-to-paramagnetic phase transition near the Curie temperature \cite{Bergqvist2018, Szilva2013} (see ${{P}_{order}}$ with the original $J$ in \textbf{Figure ~\ref{fig:F7}a}). One explanation is that the simulation of atomic spins is performed within a classical framework and ignores quantum effects \cite{Turney2009, Szilva2013}. However, as discussed below and well established in Ref. \cite{Turney2009}, quantum effects can be ignored for phonon and magnon thermal transport properties for BCC iron in the temperature range considered here. Another explanation for the transition is that the value of the magnetic exchange parameter $J$ varies as a function of temperature \cite{Bergqvist2018, Ruban2016}. 

Our EMD simulations (black and blue symbols) in \textbf{Figure ~\ref{fig:F7}a} use a single, temperature-invariant value of $J$ and treat the spins as classical, and therefore the results do not exhibit the experimentally observed sharp fall in thermal conductivity. To explore this issue further, an additional set of computations was performed (\textbf{Figure ~\ref{fig:F7}b}) wherein the value of $J$ varies with temperature. Here, we adjust the $J$ to make sure the spin in the system in the paramagnetic state is fully disordered because $J$ is only parameter that determines the spin configuration in the system. We assume that $J$ is constant when $T<{{T}_{c}}$, while $J$ is one tenth of the low temperature value when $T>{{T}_{c}}$. Thus, ${{P}_{order}}$ becomes zero when $T>{{T}_{c}}$ (\textbf{Figure ~\ref{fig:F7}b}), indicating that the spin configuration in the system is fully disordered (i.e., the paramagnetic limit). Consequently, the non-electronic thermal conductivity drops sharply around the Curie temperature due to the decrease in magnon thermal conductivity (\textbf{Figure ~\ref{fig:F7}c}), consistent with the experimental observations \cite{Fulkerson1966, Backlund1961}. 

At the same time, phonon-magnon scattering does not exhibit a strong effect on thermal transport in BCC iron because the values of both the phonon and the magnon thermal conductivities are relatively small, i.e., phonon-phonon scattering and magnon-magnon scattering are the dominant scattering mechanisms in such materials.

\subsection{\label{sec:level2}B. Phonon-magnon Scattering Process}
To further understand the temperature behavior of thermal conductivity, mode-level phonon and magnon scattering rates have been calculated at three temperatures, 300 K, 700 K and 1100 K, by spectral energy density analysis. Using Matthiessen\rq s rule, phonon-magnon ($\Gamma _{phonon-magnon}^{{}}$, in units of $1/ps$) and magnon-phonon ($\Gamma _{magnon-phonon}^{{}}$) scattering rates can be calculated via: 
\begin{equation}
\Gamma _{phonon-magnon}^{{}}=\Gamma _{phonon}^{magnon}-\Gamma _{phonon}^{no\ magnon}
\label{eqn:A24}  
\end{equation}
\begin{equation}
\Gamma _{magnon-phonon}^{{}}=\Gamma _{magnon}^{phonon}-\Gamma _{magnon}^{no\ phonon}
\label{eqn:A25}  
\end{equation}
where the superscripts $magnon$ and $no~ magnon$ indicate that the lattice vibrates with and without spins in the system, respectively, and vice versa for $phonon$ and $no~ phonon$. 

\begin{figure}[H]
\includegraphics [width=6.5in]{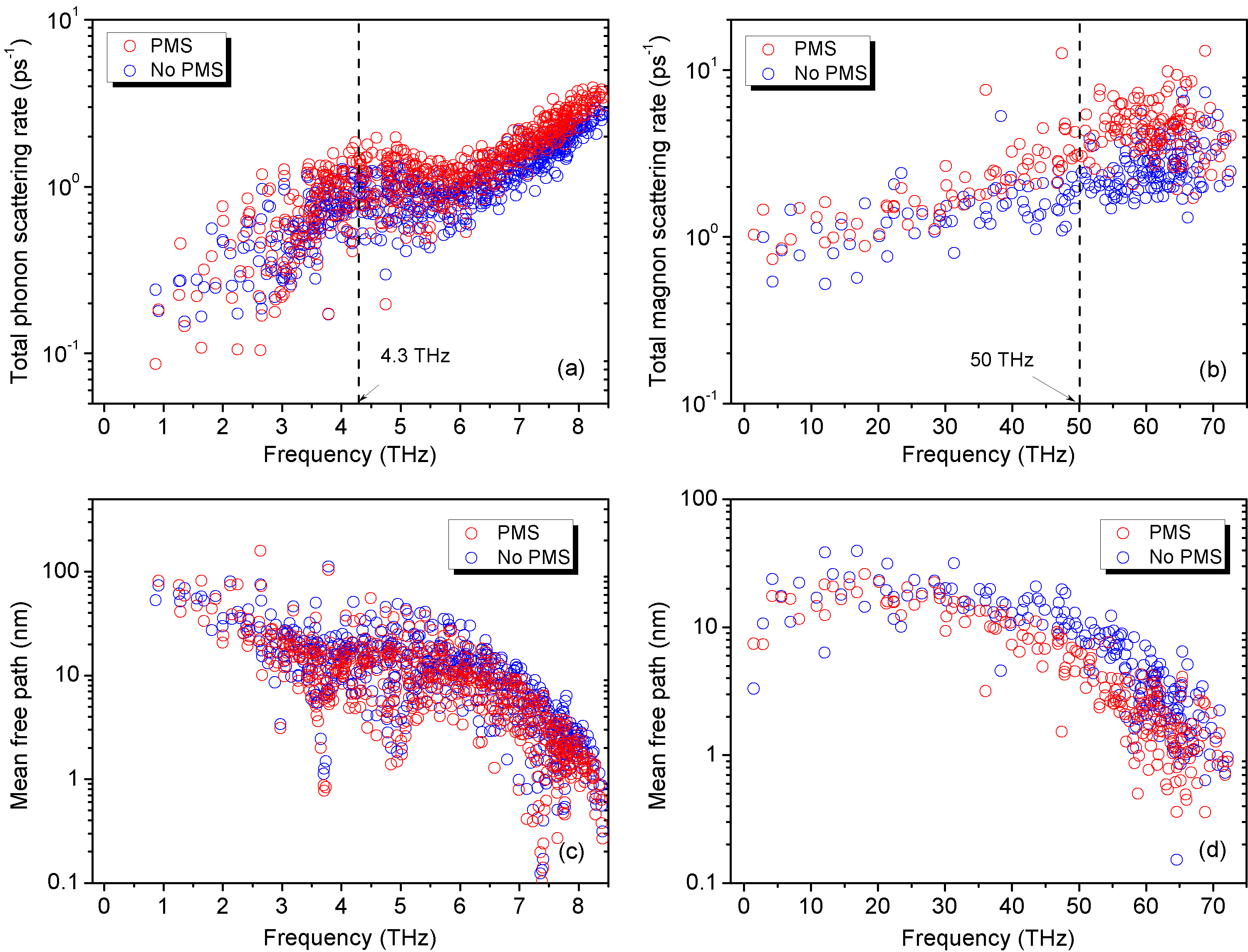}
\caption{(a) Phonon and (b) magnon scattering rate with and without considering phonon-magnon scattering at 300 K. The corresponding (c) phonon and (d) magnon mean free paths. PMS stands for the phonon-magnon scattering.}
\label{fig:F8}
\end{figure}

\begin{figure}[H]
\setlength{\abovecaptionskip}{0.1in}
\setlength{\belowcaptionskip}{-0.1in}
\includegraphics [width=6.5in]{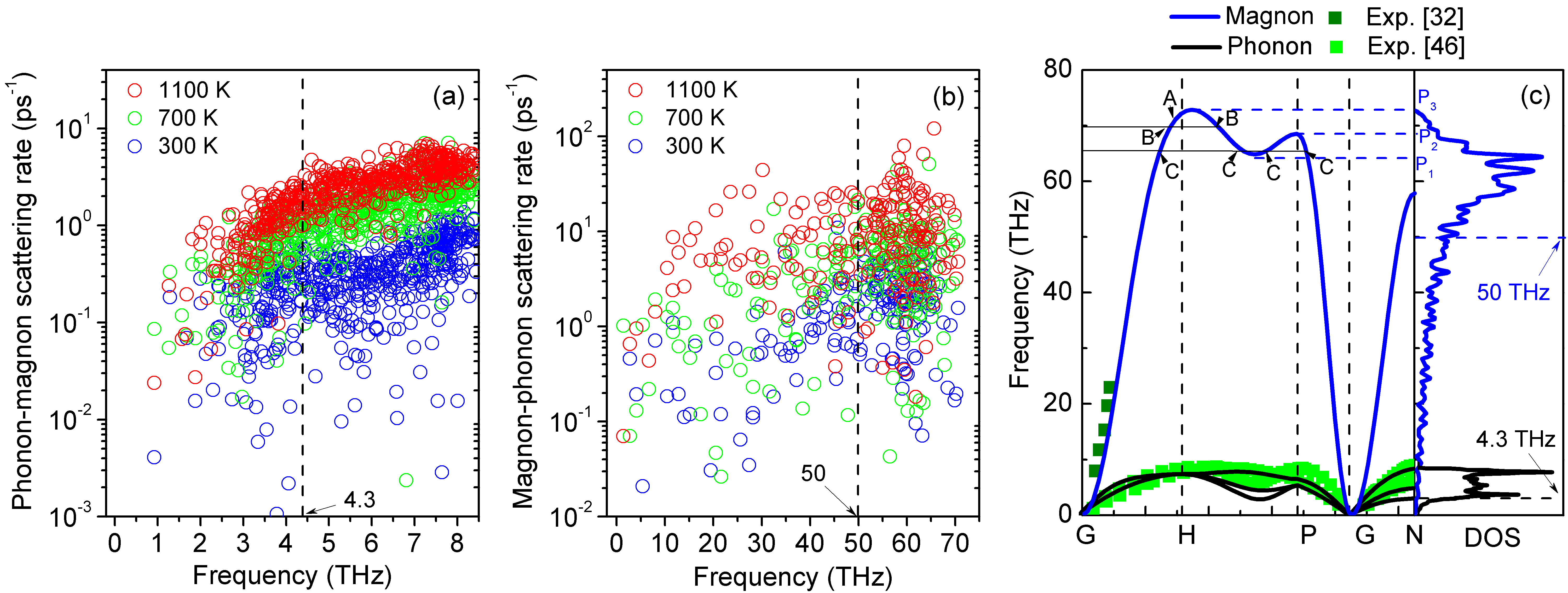}
\caption{(a) Computed phonon-magnon and (d) magnon-phonon scattering rates based on Matthiessen\rq s rule. (c) Phonon and magnon dispersions. Experimental data are from Ref. \cite{Mook1973, Minkiewicz1967}}
\label{fig:F9}
\end{figure}

First, the phonon (magnon) scattering rate of the same system with (without) considering the effects of magnons (phonons) was calculated based on spectral energy density. For phonons, our results indicate that (\textbf{Figure ~\ref{fig:F8}a}) the scattering rate in the low-frequency region changes little with the introduction of spin. Low-frequency phonons are known to be the main contributors to thermal conductivity at low temperatures whereas high-frequency phonons (i.e., short mean free path phonons) are important for lattice thermal conductivity at high temperatures \cite{Cuffe2015}. We explain why the lattice thermal conductivity considering spin is similar to that without it at low temperatures (see \textbf{Figure ~\ref{fig:F9}a}). For magnons (\textbf{Figure ~\ref{fig:F8}b}), the scattering rates for the systems with and without lattice vibrations are similar in this regime (below 50 THz), which indicates that the influence of phonon-magnon scattering on magnon transport in the low-frequency region is not important. On the other hand, in the high-frequency regime (above 50 THz), magnons can be strongly scattered by phonons.

From the phonon-magnon scattering results (\textbf{Figure ~\ref{fig:F9}a}), we observe a general tendency that high-frequency phonons ($\omega$ larger than 4.3 THz) are scattered by magnons more strongly than low-frequency phonons because the magnon energy is much larger than the phonon energy (\textbf{Figure ~\ref{fig:F9}c}). For phonon-dominant magnetic materials in which the thermal conductivity is mainly due to phonons, magnons may scatter phonons \cite{Chernyshev2015, Stamokostas2017} through phonon emission or absorption, $\hbar \omega ({\vec{k}}',\ {\mu }')=\hbar \omega (\vec{q},\ \nu )+\hbar \omega (\vec{k},\ \mu )$. Referring to \textbf{Figure ~\ref{fig:F9}c}, and considering a magnon of energy $\hbar {{\omega }_{A}}$ scattered by a magnon of energy $\hbar {{\omega }_{B}}$, a high-frequency phonon would have a greater probability of involvement in a phonon-magnon scattering process than a low-frequency phonon. For instance, only two channels exist for the magnon $\hbar {{\omega }_{A}}$ to be scattered to the magnon $\hbar {{\omega }_{B}}$ when the phonon energy is $\hbar \omega =12.41$ meV ($\omega =3$ THz), whereas the number of channels for the magnon $\hbar {{\omega }_{A}}$ to be scattered to the magnon $\hbar {{\omega }_{C}}$ is four when the phonon energy is 37.6 meV ($\omega =9.1$ THz, the highest phonon frequency). From the magnon dispersion curve, the frequency changes for the small (P1-P2) and large peaks (P2-P3) are 4.3 and 3.6 THz, respectively; therefore, phonons (magnons) above 4.3 (68) THz have a much greater chance to be involved in phonon-magnon scattering processes. 

Another reason that high-frequency phonons are preferentially involved in phonon-magnon scattering processes is the high density of states of phonons in the high-frequency region (\textbf{Figure ~\ref{fig:F9}c}). First-principles calculations \cite{Kormann2014} and experimental measurements \cite{Mauger2014} also confirm that high-frequency phonons are more strongly scattered by magnons. For materials in which the magnon thermal conductivity is dominant, magnons are scattered by phonons \cite{Chernyshev2015, Stamokostas2017} via phonon absorption or emission processes $\hbar \omega ({\vec{k}}',\ {\mu }')\pm \hbar \omega (\vec{q},\ \nu )=\hbar \omega (\vec{k},\ \mu )$. The magnon-phonon scattering rate of high-frequency magnons is somewhat higher than that of low-frequency magnons (\textbf{Figure ~\ref{fig:F9}b}) because high-frequency magnons have a larger density of states (\textbf{Figure ~\ref{fig:F9}c}). The phenomena discussed above become more apparent at increased temperatures, which strengthen the scattering among the heat carriers.

\subsection{\label{sec:level2}C. Accumulated Thermal Conductivity}
\begin{figure}[H]
\begin{center}
\setlength{\abovecaptionskip}{0.1in}
\setlength{\belowcaptionskip}{-0.1in}
\includegraphics [width=5.5in]{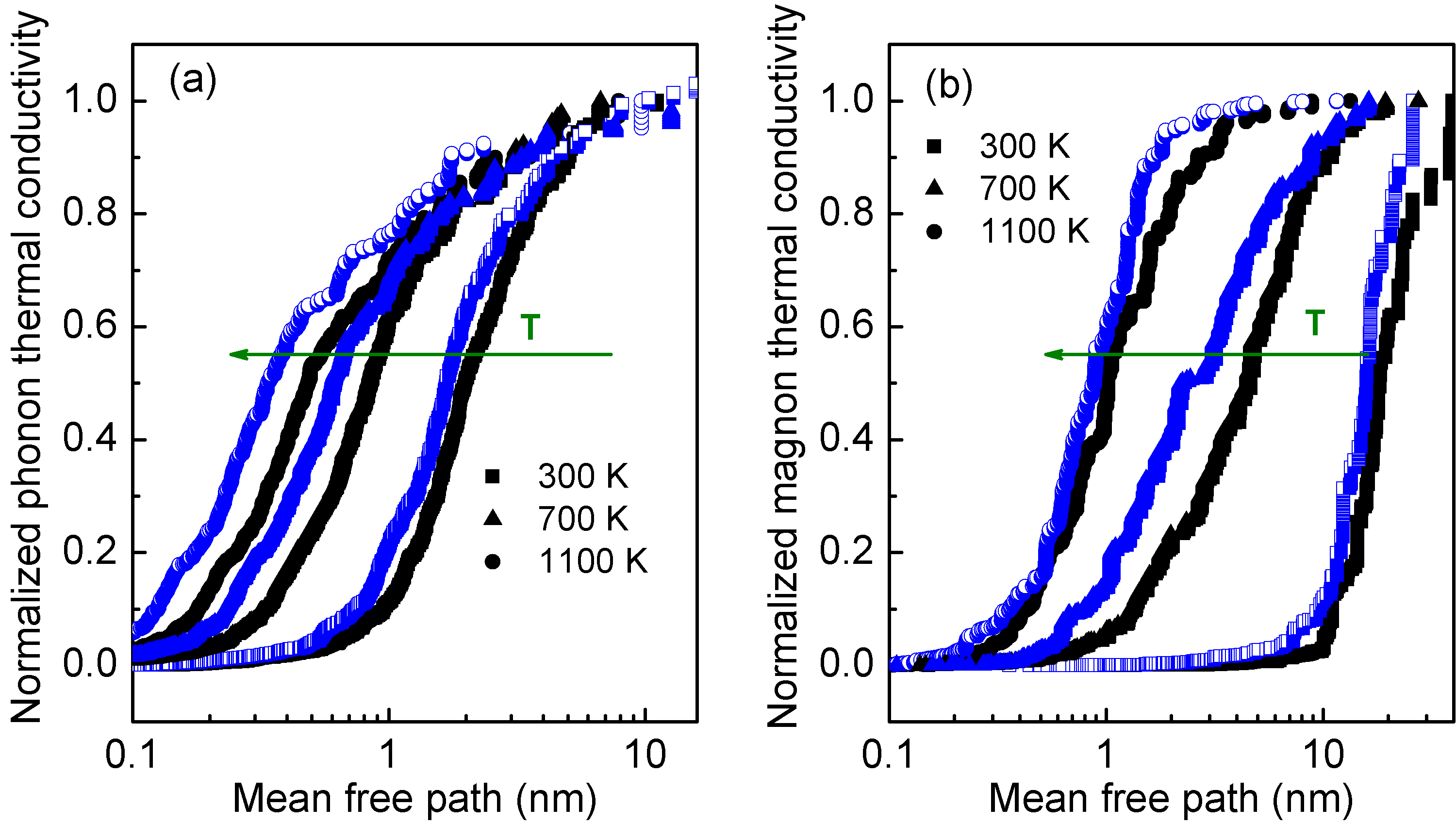}
\caption{(a) Thermal conductivity accumulation function of phonons and (b) magnons at various temperatures. The black and blue dots represent results (a) without and with spin, and (b) without and with phonons, respectively.}
\label{fig:F10}
\end{center}
\end{figure}
To facilitate the analysis of thermal conductivity contributions, the thermal conductivity accumulation function has been computed with respect to the mean free path $\Lambda$ using
\begin{equation}
\kappa (\Lambda )=\sum\limits_{\Lambda <{{\Lambda }_{0}}}{{{c}_{v}}v_{g}^{{}}\Lambda }=\sum\limits_{\Lambda <{{\Lambda }_{0}}}{{{c}_{v}}v_{g}^{2}\tau }
\label{eqn:A23}  
\end{equation}
where ${{c}_{v}}$, ${{v}_{g}}$ (\textbf{Figure ~\ref{fig:F5}}) and $\tau$ are the volumetric heat capacity, group velocity, and relaxation times of phonons and magnons, respectively. For phonons (\textbf{Figure ~\ref{fig:F10}a}), at $T$ = 300 K, the accumulated thermal conductivities with and without spin are similar, indicating that magnons do not have a strong effect on phonon thermal conductivity near room temperature. This is due to the fact that long mean free path phonons are the main contributors to thermal conductivity, and these phonons do not easily scatter with magnons, as discussed above. However, at 700 K, the phonon mean free path falls to 0.3 - 4 nm, which is much smaller than that at 300 K due to strong phonon-phonon scattering (black triangle symbols in \textbf{Figure ~\ref{fig:F10}a}). Furthermore, high-frequency phonons with short mean free paths (0.4 - 0.8 nm) are more easily scattered by magnons. Consequently, the total lattice thermal conductivity of magnetic BCC iron becomes slightly smaller than that of non-magnetic BCC iron (\textbf{Figure ~\ref{fig:F8}}). Finally, at 1100 K phonons are scattered strongly by both phonons and magnons because high temperatures increase the vibration magnitude of lattice and spin excitations. 

For magnons (\textbf{Figure ~\ref{fig:F10}b}) at room temperature, only the long-mean-free-path, or equivalently, low-frequency magnons are scattered by phonons because only these magnons transport significant energy, i.e., the heat capacity of high frequency magnons is very small (\textbf{Figure ~\ref{fig:F5}a}). When the system temperature increases to 700 K and then to 1100 K, the mean free path of magnons becomes much shorter than that at 300K, and the short mean free path (high frequency) magnons are scattered strongly by phonons. 

\section{\label{sec:level1}V. Discussion}
We also note that the magnon group velocities are calculated here from the magnon dispersion without considering temperature effects, i.e., assuming a ferromagnetic system. Such an assumption may introduce inaccuracies in the magnon mean free path and modal thermal conductivity computations. At 300 K, the total magnon thermal conductivity calculated using the Boltzmann transport equation, i.e., using  $\kappa =\sum{{{c}_{v}}v_{g}^{2}\tau }$ with the values of relaxation time computed using spectral energy density, is 10.1 W/mK. This value is lower than the Boltzmann transport equation results of Wu $et.$ $al.$ (15.2 W/mK ) due to their overestimation of the magnon dispersion \cite{Wu2018}, whereas it is 7.8 W/mK computed by GKEASD. The closer correspondence between Boltzmann transport equation and GKEASD at room temperature indicates that the system may be treated as ferromagnetic at 300 K. However, the total magnon thermal conductivities calculated by the Boltzmann transport equation (GKEASD) are 11.1 (1.9) and 56.9 (1.2) W/mK at 700 K and 1100 K, respectively. The large differences between spectral energy density-Boltzmann transport equation and GKEASD indicate that the magnon group velocity is overestimated, and the calculation should consider spin disorder at high temperatures.

Before concluding our study, we briefly highlight two limitations that our methodology inherited from the spin lattice dynamics formalism. In our calculations, the exchange integral $J_{ij}$ is assumed to remain constant, and its dependence on temperature is neglected. Recent studies have investigated this dependence and proposed methodologies to account for it \cite{Szilva2013, Ruban2016}. Encapsulating them within our framework could improve the accuracy of predictions. Furthermore, the simulation of classical spins (instead of quantum spins) is known to make the ferromagnetic to paramagnetic phase transition smoother than in the experimental observations. It has been shown in \cite{Bergqvist2018, Szilva2013} that implementing quantum baths and statistics can reproduce more accurately the sharp transitin at $T_c$. However, as the GKESAD framework is based on equilibrium molecular dynamics, the associated simulations do not involve a connection to a random bath. Overall, an empirical parametrization of the exchange integral $J_{ij}$ could account for both its temperature dependence and the sharp transition occuring at $T_c$. 

\section{\label{sec:level1}VI. Conclusions}
In conclusion, we have developed a temperature-dependent method, the Green-Kubo equilibrium atomic and spin dynamics method, to calculate coupled phonon and magnon transport in magnetic materials. Reasonably good agreement is obtained between our simulation results and experimental measurements in computing dispersion curves and temperature-dependent thermal conductivity, and these results suggest that the approach captures the overall heat transfer behavior of phonons and magnons in magnetic crystals. However, the coupling between is governed by the magnetic exchange correlation constant, whose origins and fidelity deserve further scrutiny, as do other magnetic phenomena not considered here, such as anisotropy \cite{Finizio2014}. Analysis of scattering processes between phonons and magnons indicates that high-frequency phonon scattering rates due to phonon-magnon scattering are much larger than those at low frequencies because of energy-conserving rules for scattering and the high density of states. The application of this new methodology will yield deeper insights into the thermal transport properties of other phonon- or magnon-dominant materials.




\section{\label{sec:level1} Acknowledgments}
This work was supported by the National Science Foundation (NSF) (Project number: NSF1758004) and used computational and storage services associated with the Hoffman2 Shared Cluster provided by UCLA Institute for Digital Research and Education\rq s Research Technology Group. This work used the Extreme Science and Engineering Discovery Environment (XSEDE), which is supported by National Science Foundation grant number DMR180057. Y. Z. gratefully acknowledges Dr. Zheyong Fan (Aalto University) for valuable comments and proof reading the manuscript. J. T. acknowledges that Sandia National Laboratories is a multi-mission laboratory managed and operated by National Technology and Engineering Solutions of Sandia, LLC, a wholly owned subsidiary of Honeywell International Inc., for the U.S. Department of Energy\rq s National Nuclear Security Administration under contract DE-NA0003525. This paper describes objective technical results and analysis. Any subjective views or opinions that might be expressed in the paper do not necessarily represent the views of the U.S. Department of Energy or the United States Government.

\end{spacing}

\end{document}